\documentclass[lettersize,journal]{IEEEtran}

\usepackage{amsmath,amsfonts,amssymb}
\usepackage{algorithmic}
\usepackage{algorithm}
\usepackage{array}
\usepackage[caption=false,font=footnotesize,labelfont=rm,textfont=rm]{subfig}
\usepackage{textcomp}
\usepackage{stfloats}
\usepackage{url}
\usepackage{verbatim}
\usepackage{graphicx}
\usepackage{cite}
\usepackage{balance}

\usepackage{booktabs}
\usepackage{multirow}

\graphicspath{{figures/}}

\newcommand{\cmark}{\ensuremath{\checkmark}}
\newcommand{\xmark}{\ensuremath{\times}}

\makeatletter
\def\section{\@startsection{section}{1}{\z@}%
{1.80ex plus 0.60ex minus 0.60ex}%
{0.35ex plus 0.15ex minus 0ex}%
{\normalfont\normalsize\centering\scshape}}%
\def\subsection{\@startsection{subsection}{2}{\z@}%
{1.60ex plus 0.50ex minus 0.50ex}%
{0.30ex plus 0.12ex minus 0ex}%
{\normalfont\normalsize\itshape}}%
\def\subsubsection{\@startsection{subsubsection}{3}{\parindent}%
{0pt}%
{0pt}%
{\normalfont\normalsize\itshape}}%
\makeatother

\AtBeginDocument{%
\setlength{\abovedisplayskip}{4pt plus 1pt minus 1pt}%
\setlength{\belowdisplayskip}{4pt plus 1pt minus 1pt}%
\setlength{\abovedisplayshortskip}{2pt plus 1pt minus 1pt}%
\setlength{\belowdisplayshortskip}{3pt plus 1pt minus 1pt}%
\setlength{\jot}{2pt}%
}

\begin{document}

\title{{\fontsize{23.5pt}{23pt}\selectfont
Service-Adaptive Policy Learning for Physical-Layer Security with Pinching-Antenna Systems}}

\author{Zhaoming Hu, Xiaochen Nie, Ruikang Zhong, Haochen Li, Dengao Li, and Xidong Mu%

\vspace{-1cm}
\thanks{Z. Hu, X. Nie, and D. Li are with the College of Computer Science and Technology (College of Data Science), Taiyuan University of Technology, Taiyuan 030024, P.R. China.}%
\thanks{Z. Hu and D. Li are also with the Key Laboratory of Data Governance and Intelligent Decision-Making of Shanxi Province, Taiyuan 030024, P.R. China.}%
\thanks{R. Zhong is with the School of Information and Communication Engineering, Xi'an Jiaotong University, Xi'an 710049, P.R. China.}%
\thanks{H. Li is with the College of Electronic and Information Engineering, Nanjing University of Aeronautics and Astronautics, Nanjing 211106, China, and also with the National Mobile Communications Research Laboratory, Southeast University, Nanjing 210096, China.}%
\thanks{X. Mu is with the Centre for Wireless Innovation (CWI), Queen's University Belfast, Belfast BT3 9DT, U.K.}%
}

\markboth{IEEE Transactions on Mobile Computing}%
{Hu \MakeLowercase{\textit{et al.}}: Security-Aware Pinching-Antenna Systems (PASS): Physical-Layer Security Transmission}

\maketitle
\begin{abstract}
Confidentiality protection in heterogeneous mobile services depends on both service-specific security requirements and stream-specific receiver authorization. A receiver authorized for one information stream may therefore be a potential interceptor for another. Leveraging the flexibility of physical-layer control to adapt confidentiality protection across receivers and information streams, we investigate a pinching-antenna system (PASS)-enabled physical-layer security (PLS) framework, in which movable pinching antennas (PAs) reshape the effective channel disparities between intended and unintended receivers. A unified transmission model is established for different security demands level, where receiver roles and performance objectives are determined by the active security requirement. Security-aware transmit beamforming and PA positions are jointly optimized to maximize the worst-user rate or secrecy rate. Considering the temporal coupling introduced by PA repositioning, the problem is formulated as a long-horizon Markov decision process. To solve the resulting MDP, we develop two policy-learning algorithms with different design priorities. Designed for lightweight and stable cross-mode control, the heterogeneous security-aware proximal policy optimization (HSPPO) algorithm evaluates sampled transitions based on mode-specific performance and constraint violations. For stronger representation and optimization capabilities, the multi-relational hierarchy-aware diffusion policy optimization (MRHA-DPO) algorithm integrates multi-relational PASS state representation with hierarchy-aware diffusion-based action generation to capture heterogeneous interactions and coordinate coupled control decisions. Simulation results show that movable PAs consistently outperform fixed position PA and conventional multiple-input multiple-output (MIMO) in rate and secrecy performance, with increasing gains at higher transmit power. Moreover, HSPPO and MRHA-DPO outperform conventional PPO and TD3 across all security modes, demonstrating favorable convergence and robust adaptability to heterogeneous security requirements.
\end{abstract}


\section{Introduction}

\IEEEPARstart{D}{riven} by the rapid development of emerging mobile applications, wireless networks are evolving from providing homogeneous data connectivity toward supporting heterogeneous services such as immersive communications, industrial automation, digital twins, and massive Internet-of-Things~(IoT) applications~\cite{Alwis2026COMST}. The coexistence of these services over a shared wireless infrastructure introduces increasingly diverse requirements for latency, reliability, throughput, and confidentiality. Focusing on confidentiality, public or non-sensitive information may prioritize transmission rate without explicit secrecy protection. Commercially sensitive data may require protection from external eavesdroppers, whereas mission-critical or privacy-sensitive information may additionally need to be protected from legitimate users that are not authorized to access the target information. Consequently, future mobile networks need service-adaptive transmission mechanisms that accommodate heterogeneous confidentiality requirements and stream-specific authorization relationships.

Physical-layer security (PLS) exploits channel disparities between intended and unintended receivers to protect wireless transmissions~\cite{Zhao2025TWC}. It complements upper-layer cryptographic mechanisms by providing confidentiality protection directly at the physical layer. Secure beamforming, artificial noise (AN), cooperative jamming, and power allocation have been widely investigated to suppress information leakage during wireless transmission~\cite{Niu2026COMST,Zhang2026TMC}. However, most existing PLS schemes assume fixed threat models and apply uniform secrecy objectives across services. For services with low confidentiality requirements, such protection may impose unnecessary secrecy constraints and reduce the achievable rate. For highly sensitive services, protecting only against external eavesdroppers overlooks potential information leakage to non-target legitimate receivers that are not authorized to decode those streams. Service-dependent PLS should therefore account for stream-specific differences in performance objectives and receiver roles.

The effectiveness of PLS largely depends on channel disparities between intended receivers and potential eavesdroppers. However, such disparities can be limited in conventional fixed-antenna systems, particularly when users are closely located or experience similar propagation conditions~\cite{Li2024JSAC}. Recently, spatially reconfigurable antenna technologies have been investigated to reshape wireless channels by introducing additional spatial degrees of freedom~\cite{New2026JSAC,Chen2026TMC}. Among these technologies, the pinching-antenna system (PASS) provides a flexible implementation for reconfiguring signal propagation. PASS transports radio-frequency signals through low-loss dielectric waveguides and radiates them into free space through pinching antennas (PAs) deployed at adjustable positions~\cite{Xu2026COMST}. By relocating the pinching points, PASS can simultaneously change the in-waveguide propagation phase and free-space transmission distance, thereby adapting the composite channel without requiring an independent radio-frequency chain for each PA~\cite{Li2026WCOM}. This capability enables PASS to enhance desired signals toward intended users while reducing signal exposure to unintended receivers.

Integrating PASS into a service-adaptive PLS framework provides an opportunity to translate heterogeneous confidentiality requirements into adaptive physical-layer transmission strategies. For service with a low security requirement, PA configurations and transmit signals can be optimized primarily to improve the worst-user rate. For services requiring protection against external eavesdropping, PASS can reshape wireless propagation to enhance legitimate links while suppressing information leakage to external eavesdroppers. For highly confidential services, authenticated yet unauthorized users should additionally be treated as potential interceptors, requiring stronger stream-level isolation. The same receiver may therefore be authorized for one information stream while acting as a potential interceptor for another. Consequently, the security level determines not only the required secrecy performance, but also the receiver roles and corresponding interceptor sets. Jointly adapting PA positions, transmit beamforming, and AN is thus essential to support heterogeneous secure transmission over a shared PASS infrastructure.

\subsection{State-of-the-Art}

The PASS-enabled security-aware PLS framework developed in this work combines propagation reconfiguration through movable PAs with learning-based joint control under heterogeneous security requirements. This integration builds on recent advances in Flexible-Antenna Technologies and learning-based secure transmission optimization.

\subsubsection{Flexible-Antenna Technologies for Physical-Layer Security}

Flexible-Antenna Technologies enhance PLS by introducing controllable spatial or electromagnetic degrees of freedom. Fluid and movable antenna-assisted schemes exploit position-dependent channel variations to strengthen legitimate reception or weaken eavesdropping links~\cite{Feng2025TCOM,Li2025TIFS,Ma2025TWC,xiao2024TWC_MA}. Ghadi \emph{et al.} analyzed secrecy capacity, secrecy outage probability, and secrecy energy efficiency of fluid-antenna systems under correlated fading~\cite{Rostami2024TWC}. Ding \emph{et al.} jointly optimized movable antenna positions, transmit and receive beamforming, AN beamforming, and uplink power for secure full-duplex multi-user communications~\cite{Ding2025TWC}. Reconfigurable intelligent surface (RIS)-assisted schemes adjust the phase responses of passive reflecting elements to favor legitimate users while suppressing signals at unintended receivers~\cite{Tishchenko2025COMST,zhu2024TGCN}. Peng \emph{et al.} jointly designed base-station beamforming and RIS phase shifts for robust secure multi-user transmission under hardware impairments~\cite{peng2023TWC}. Xiao \emph{et al.} jointly optimized the active and passive beamformers of a simultaneously transmitting and reflecting RIS-assisted millimeter-wave system for PLS and covert communication~\cite{xiao2024TWC}. Unlike antenna relocation or passive reflection, PASS jointly controls guided-wave propagation and free-space radiation through adjustable pinching points.

In PASS-assisted systems, existing studies have jointly optimized PA placement and transmit beamforming to control guided-wave phases and free-space propagation distances~\cite{Shan2026TCOM,Xu2026LWC}. Zhu \emph{et al.} developed secure PASS architectures that coordinate pinching beamforming, baseband beamforming, AN transmission, and power allocation~\cite{Zhu2026TCOM}. Jiang \emph{et al.} optimized PA positions and transmit beamforming to improve covert communication~\cite{jiang2026JSAC}, while Zhao \emph{et al.} jointly designed beamforming, AN covariance, PA power ratios, and PA positions under blockage and eavesdropper-channel uncertainty~\cite{zhao2026TMC}. These studies demonstrate that guided-wave propagation, PA positioning, transmit beamforming, and AN can be coordinated to strengthen legitimate links and suppress information leakage~\cite{Li2026TMC}. However, current secure PASS schemes mainly assume a fixed security topology and optimize a common secrecy or covertness objective. They do not consider service-dependent authorization, under which the same legitimate receiver may be authorized for one stream but act as a potential interceptor of another. Consequently, supporting heterogeneous services requires PASS configurations to adapt to both the physical network topology and the stream-dependent security topology.

\subsubsection{Learning-Based Control for Secure Wireless Transmission}

Learning-based methods have been widely investigated to address the computational complexity and dynamic nature of secure wireless transmission~\cite{Zhong2026TCCN,Waqar2024TWC,Zhong2024MNET}. Deep reinforcement learning (DRL) can learn transmission policies from time-varying channel states, user locations, and adversarial conditions, avoiding repeated solutions of non-convex optimization problems~\cite{Feng2026TWC,wang2024TWC_FAS}. Beyond secure transmission, learning-driven methods have also been extensively studied for intelligent control and resource optimization in mobile networks~\cite{Cao2026TMC,Wu2026TMC,Liang2026TMC}. Wang \emph{et al.} combined double deep Q-learning and soft actor--critic learning to jointly optimize RIS pairing, subchannel assignment, transmit beamforming, reflection coefficients, and sensing time in a secure multi-RIS-assisted spectrum-sharing network~\cite{wang2024TWC}. For sensing-aided covert PASS transmission, Jiang \emph{et al.} used DRL to adapt PA positions to the correlated mobility of a tracked warden while jointly designing transmit beamforming and AN~\cite{jiang2026JSAC_Sensing}. Graph neural network (GNN)-based methods represent legitimate users, eavesdroppers, and wireless links as entities to capture topology-dependent interference and security relationships~\cite{zhang2026JSAC,zhang2026JIOT}. Zhang \emph{et al.} developed a GNN-based secrecy-rate optimization framework for multi-satellite collaborative systems involving satellites, legitimate users, and eavesdroppers~\cite{zhang2026JSAC}. Zhang \emph{et al.} also developed a GNN-based framework for RIS-assisted integrated sensing and communication (ISAC) systems, exploiting graph representations of users and eavesdroppers to optimize secure transmission and RIS configuration for improved secrecy performance~\cite{zhang2026JIOT}. Nevertheless, these learning-based secure-transmission methods assume predefined receiver roles and are tailored to reconfigurable architectures or threat models.

Recent PASS studies have introduced structure-aware neural models for coupled PA-positioning and beamforming decisions~\cite{xu2026TWC,guo2026JSAC}. Mu \emph{et al.} developed a Karush--Kuhn--Tucker (KKT)-guided dual-learning Transformer capturing inter-PA, inter-user, and channel state information (CSI)--beamforming dependencies to generate transmit and pinching beamforming decisions~\cite{xu2026TWC}. Guo \emph{et al.} proposed PASSformer, a graph Transformer with permutation properties, to coordinate digital, analog, and pinching beamforming across varying numbers of users, waveguides, and PAs~\cite{guo2026JSAC}. Although these methods provide structural representations for coupled PASS control, they focus on communication rate or spectral efficiency rather than secure transmission. Moreover, existing learning-based PLS and PASS methods do not jointly account for the physical PA--user topology and stream-dependent security relationships, nor do they explicitly distinguish the type and severity of performance violations across security modes.
\subsection{Motivation and Challenges}

Heterogeneous services with different security requirements may coexist in the same wireless network and share a common PASS infrastructure. Designing an independent transmission strategy for each security mode would increase control complexity and limit knowledge sharing across services. A unified framework that adapts PASS configurations to different protection requirements is therefore desirable. However, such unification presents the following challenges.

First, heterogeneous security requirements lead to different receiver roles and performance criteria. Low-security transmission (LST) primarily concerns reliable delivery, medium-security transmission (MST) protects each stream from external eavesdroppers, whereas high-security transmission (HST) must additionally prevent non-target legitimate users from decoding it. These modes differ in intended receivers, potential interceptors, and achievable-rate metrics, making unified optimization over the same PASS architecture difficult.

Second, mode-specific optimization makes training samples unequally informative to a shared DRL policy. Samples from different security modes reflect distinct objectives, constraint violations, and performance scales. Treating them equally during policy updates may cause certain modes to dominate learning, whereas separate policies prevent knowledge sharing across modes. Effectively utilizing mode-dependent samples in a unified framework is therefore challenging.

Third, PASS-specific physical and security relationships complicate both state representation and joint action generation. PA--user propagation, stream-dependent receiver roles, and inter-PA placement constraints create distinct dependencies within the system state. Meanwhile, information-bearing beamforming, AN beamforming, and PA positioning operate in different domains but jointly determine transmission and secrecy performance. Flattened state representations may obscure these dependencies, while conventional unimodal Gaussian policies may inadequately model the coupled action space.

\subsection{Contributions}

To support adaptive secure transmission under heterogeneous confidentiality requirements, we establish a PASS-enabled security-aware PLS framework and develop two complementary learning-based controllers with different complexity–performance tradeoffs. The contributions are as follows:

\begin{itemize}
    \setlength{\topsep}{2pt}
    \setlength{\itemsep}{0pt}
    \setlength{\parsep}{0pt}
    \setlength{\parskip}{0pt}
    \item We establish a PASS-enabled security-aware PLS framework for low-, medium-, and high-security transmission, where receiver roles and performance objectives are determined by the active security requirement. Security-aware transmit beamforming and PA positioning are jointly optimized in a long-horizon control problem that accounts for PA repositioning.

    \item We develop a heterogeneous security-aware proximal policy optimization (HSPPO) algorithm for lightweight and stable cross-mode control. By incorporating mode-specific performance and constraint violations into policy updates, HSPPO evaluates each sampled transition according to its active security mode. This enables mode-specific updates within a unified policy and improves the utilization of heterogeneous training samples.

    \item We propose a multi-relational hierarchy-aware diffusion policy optimization (MRHA-DPO) algorithm for stronger representation and joint-optimization capabilities. MRHA-DPO captures PASS-specific physical and security relationships through multi-relational state representation and models the structural differences and coupling among information-bearing beamforming, AN beamforming, and PA-positioning decisions through hierarchy-aware diffusion-based action generation.
\end{itemize}



\section{System Model and Problem Formulation}

\begin{figure}[!t]
\centering
\includegraphics[width=\columnwidth]{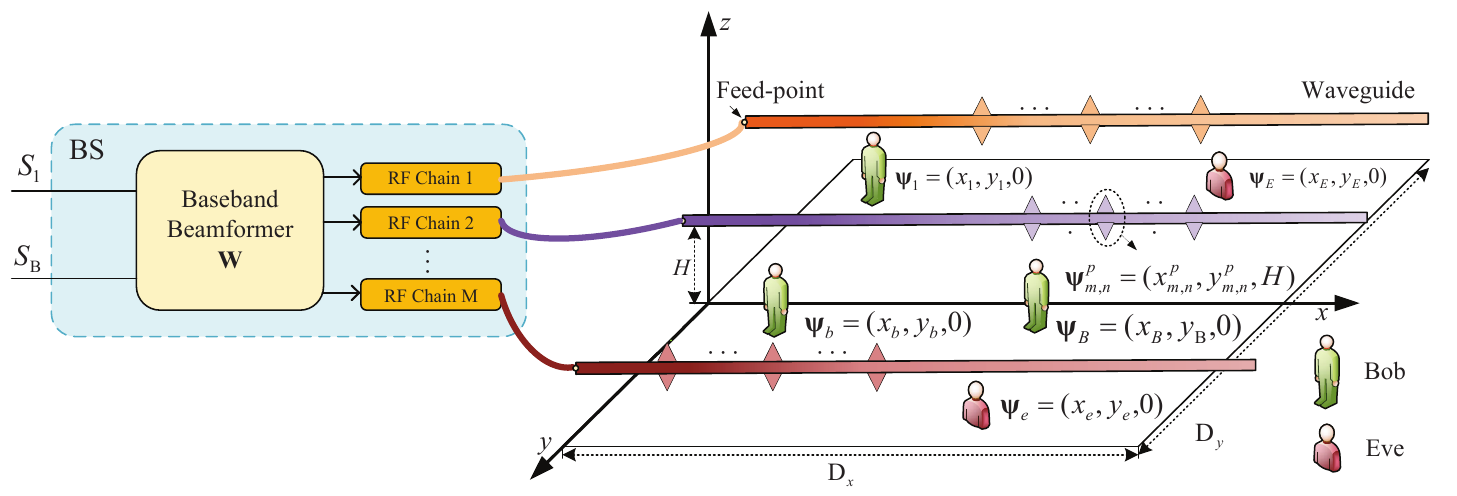}
\caption{PASS-enabled secure-aware downlink PLS system.}
\label{fig:pass_system}
\end{figure}

As illustrated in Fig.~\ref{fig:pass_system}, we consider a PASS-enabled downlink secure communication system, where a base station (BS) equipped with $N$ parallel dielectric waveguides serves multiple legitimate users in the presence of external eavesdroppers. All waveguides are deployed at a fixed height $H$ above a rectangular service region of size $D_x\times D_y$, in which all legitimate users and potential eavesdroppers are located. Let $\mathcal N\triangleq\{1,\ldots,N\}$ and $\mathcal M\triangleq\{1,\ldots,M\}$ denote the index sets of waveguides and PAs on each waveguide, respectively. The $N$ waveguides are uniformly arranged along the $y$-axis with inter-waveguide spacing $d_y=D_y/(N-1)$, and each waveguide is oriented along the $x$-axis. The feed point and the $m$-th PA on the $n$-th waveguide are located at $\boldsymbol\psi_{n,0}^{p}=[0,\,y_n,\,H]^{\mathsf T}$ and $\boldsymbol\psi_{n,m}^{p}=[x_{n,m},\,y_{n,m},\,h]^{\mathsf T}$, respectively. The PA positions on each waveguide satisfy $0\leq x_{n,1}<\cdots<x_{n,M}\leq D_x$. To mitigate mutual coupling effects, a minimum inter-antenna spacing is imposed as $x_{n,m+1}-x_{n,m}\geq\Delta_{\min}=\lambda/2$, $\forall m\leq M-1$,
where $\lambda$ denotes the free-space carrier wavelength. All terminals are partitioned into two disjoint sets, namely the legitimate user (Bob) set $\mathcal B$ and the eavesdropper (Eve) set $\mathcal E$. Let $\mathcal K=\mathcal B\cup\mathcal E$ denote the set of all users. The position of user $k\in\mathcal K$ is given by $\boldsymbol\psi_k=[x_k,\,y_k,\,0]^{\mathsf T}$.

\subsection{System Model}

In the considered system, only the legitimate users request confidential information. The confidential symbols intended for the Bobs are first linearly precoded at the BS, upconverted to the carrier frequency, and then injected into the waveguides before being radiated into free space by the activated PAs. Accordingly, the aggregate transmitted signal is expressed as
\begin{equation}
\mathbf x
=
\mathbf G
\left(
\sum\nolimits_{b\in\mathcal B}\mathbf w_b s_b
+
\mathbf w_{\mathrm{AN}}z
\right),
\label{eq:aggregate_signal}
\end{equation}
where $s_b\sim\mathcal{CN}(0,1)$ denotes the confidential symbol intended for Bob $b$, $\mathbf w_b\in\mathbb C^{N\times1}$ is the transmit beamforming vector for Bob $b$, $\mathbf w_{\mathrm{AN}}\in\mathbb C^{N\times1}$ is the AN beamforming vector, and $z\sim\mathcal{CN}(0,1)$ is the AN symbol. The in-waveguide propagation is characterized by the security-aware PASS beamforming matrix $\mathbf G$. Due to the physical independence among waveguides, it is modeled as
\begin{equation}
\mathbf G
=
\operatorname{BlkDiag}\!\left(
\mathbf g_1,\mathbf g_2,\ldots,\mathbf g_N
\right),
\label{eq:pass_beamforming_matrix}
\end{equation}
where $\mathbf g_n\in\mathbb C^{M\times 1}$ denotes the
guided-wave response vector of the $n$-th waveguide. The
coefficient associated with its $m$-th PA is given by
\begin{equation}
[\mathbf g_n]_m
=
\frac{1}{\sqrt{M}}
\exp\!\left(
-jk_g
\left\|
\boldsymbol{\psi}_{n,m}^{p}
-
\boldsymbol{\psi}_{n,0}^{p}
\right\|_2
\right),
\label{eq:guided_wave_response}
\end{equation}
where $k_g=2\pi/\lambda_g$ is the guided wavenumber and
$\lambda_g=\lambda/n_{\mathrm{eff}}$ is the guided wavelength,
with $n_{\mathrm{eff}}$ denoting the effective refractive index.
We adopt an ideal equal-power radiation model, under which the
signal power carried by each waveguide is uniformly allocated
among its $M$ PAs. Accordingly, the factor $1/\sqrt{M}$
normalizes the guided-wave response such that
$\|\mathbf g_n\|_2^2=1$. The composite channel from the PASS aperture to receiver $k$ is written as
\begin{equation}
\mathbf h_k^{\mathsf H}
=
\left[
\mathbf h_{k,1},\ldots,\mathbf h_{k,n},\ldots,\mathbf h_{k,N}
\right],
\label{eq:composite_channel}
\end{equation}
which captures the superposition of signals radiated from all activated PAs over the PASS aperture. Specifically, assuming a line-of-sight (LoS) propagation model, the channel coefficient from the $m$-th PA on the $n$-th waveguide to receiver $k$ is given by
\begin{equation}
[\mathbf h_{k,n}]_m
=
\frac{
\sqrt{\eta}\exp\!\left(
-jk_0\left\|\boldsymbol\psi_{n,m}^{p}-\boldsymbol\psi_k\right\|
\right)
}{
\left\|\boldsymbol\psi_{n,m}^{p}-\boldsymbol\psi_k\right\|
},
\label{eq:subchannel_vector}
\end{equation}
with $k_0=2\pi/\lambda$ and $\eta=(\lambda/4\pi)^2$ denoting the free-space path-loss constant. Accordingly, the effective channel for receiver $k$ is defined as
\begin{equation}
\mathbf H_k^{\mathsf H}
=
\mathbf h_k^{\mathsf H}\mathbf G.
\label{eq:effective_channel}
\end{equation}
The received signal at receiver $k$ is expressed as
\begin{equation}
\begin{aligned}
y_k
&=
\mathbf h_k^{\mathsf H}\mathbf x+z_k\\
&=
\mathbf h_k^{\mathsf H}\mathbf G
\left(
\sum\nolimits_{i\in\mathcal B}\mathbf w_i s_i
+
\mathbf w_{\mathrm{AN}}z
\right)
+z_k,
\end{aligned}
\label{eq:received_signal}
\end{equation}
where $z_k\sim\mathcal{CN}(0,\sigma_k^2)$ denotes the additive
noise at receiver~$k$.

\subsection{Security-Aware PASS Transmission Model}

Legitimate users and eavesdroppers have different decoding objectives. Specifically, each legitimate user aims to decode its own intended signal, while each eavesdropper attempts to intercept the signal intended for a specific user. Accordingly, for a legitimate Bob $b\in\mathcal B$, the received signal-to-interference-plus-noise ratio (SINR) is given by
\begin{equation}
\mathrm{SINR}_b
=
\frac{
\left|\mathbf h_b^{\mathsf H}\mathbf G\mathbf w_b\right|^2
}{
\displaystyle\sum\nolimits_{\substack{i\in\mathcal B\\i\neq b}}
\left|\mathbf h_b^{\mathsf H}\mathbf G\mathbf w_i\right|^2
+
\left|\mathbf h_b^{\mathsf H}\mathbf G\mathbf w_{\mathrm{AN}}\right|^2
+
\sigma_b^2
},
\label{eq:bob_sinr}
\end{equation}
and the corresponding achievable rate is
\begin{equation}
R_b
=
\log_2\!\left(1+\mathrm{SINR}_b\right).
\label{eq:bob_rate}
\end{equation}

However, for an eavesdropper $e\in\mathcal E$ attempting to decode the signal intended for Bob $b$, the SINR is defined with respect to Bob $b$, reflecting its target-dependent interception capability. The received $\mathrm{SINR}_{e,b}$ for Eve $e$ eavesdropping on Bob $b$ is given by
\begin{equation}
\mathrm{SINR}_{e,b}
=
\frac{
\left|\mathbf h_e^{\mathsf H}\mathbf G\mathbf w_b\right|^2
}{
\displaystyle\sum\nolimits_{\substack{i\in\mathcal B\\i\neq b}}
\left|\mathbf h_e^{\mathsf H}\mathbf G\mathbf w_i\right|^2
+
\left|\mathbf h_e^{\mathsf H}\mathbf G\mathbf w_{\mathrm{AN}}\right|^2
+
\sigma_e^2
},
\label{eq:eve_sinr}
\end{equation}
and the corresponding achievable eavesdropping rate of Eve $e$ when intercepting the signal intended for Bob $b$ is
\begin{equation}
R_{e,b}
=
\log_2\!\left(1+\mathrm{SINR}_{e,b}\right).
\label{eq:eve_rate}
\end{equation}

To capture heterogeneous security requirements across different services, we classify user demands into three security levels. For low-security services, confidentiality is not strictly required, and all other Bobs and Eves are treated equivalently as legitimate receivers. For medium-security services, confidentiality is required only against Eves, while other Bobs are not considered adversarial. For high-security services, strict confidentiality is imposed, where both other Bobs and Eves are treated as potential interceptors. This classification provides a unified framework to characterize security-aware transmission design in PASS-assisted systems. Based on the above classification, we formulate three security-aware PASS transmission modes corresponding to low, medium, and high confidentiality requirements.

\subsubsection{Low-Security Transmission (LST)}

For services with low confidentiality requirements, no explicit
secrecy constraint is imposed, and the other receivers are not
regarded as potential interceptors. The transmission design
therefore focuses on improving the achievable rates of the
intended Bobs without explicitly suppressing information leakage.
Since multiple Bobs are served simultaneously, their individual
rates may differ due to heterogeneous channel conditions.
To ensure inter-user fairness, the LST performance metric is
defined as the worst-user achievable rate
\begin{equation}
R^{\mathrm{LST}}
\triangleq
\min_{b\in\mathcal B}R_b.
\label{eq:lst_metric}
\end{equation}
Maximizing $R^{\mathrm{LST}}$ prevents the transmission strategy
from favoring Bobs with stronger channels at the expense of
poorly served users.
\subsubsection{Medium-Security Transmission (MST)}

For medium-security services, confidentiality is required against
external Eves, while non-target Bobs are not regarded as potential
interceptors. Compared with LST, the transmission design must
not only maintain the desired rates of the intended Bobs but also
suppress information leakage to external Eves. Since multiple
Eves may attempt to intercept the stream intended for Bob $b$,
the maximum eavesdropping rate is used to characterize the
worst-case external leakage. Accordingly, the MST performance
metric is defined as the worst-user secrecy rate
\begin{equation}
R^{\mathrm{MST}}
\triangleq
\min_{b\in\mathcal B}
\left[
R_b-\max_{e\in\mathcal E}R_{e,b}
\right]^+.
\label{eq:mst_metric}
\end{equation}
This metric jointly accounts for fairness among the Bobs and
secrecy protection against the most capable external Eve.
\subsubsection{High-Security Transmission (HST)}

For high-security services, confidentiality is required against
all unintended receivers, including both external Eves and
non-target Bobs. Although a non-target Bob is a legitimate
network user, it may not be authorized to access the information
stream intended for Bob $b$ and is therefore treated as a
potential interceptor. Compared with MST, HST thus requires
stronger stream-level isolation by extending the potential-
interceptor set to all receivers except the intended Bob.
Accordingly, the HST performance metric is defined as
\begin{equation}
R^{\mathrm{HST}}
\triangleq
\min_{b\in\mathcal B}
\left[
R_b-\max_{i\in\mathcal K\setminus\{b\}}R_{i,b}
\right]^+,
\label{eq:hst_metric}
\end{equation}
where $R_{i,b}$ denotes the achievable rate at receiver $i$
when attempting to decode the information stream intended for
Bob $b$, following the same formulation as $R_{e,b}$.
By considering the most capable unintended receiver, this metric
provides the strictest confidentiality protection among the three
transmission modes.

\subsection{Problem Formulation}
\label{sec:problem_formulation}

The proposed security-aware PASS beamforming strategy operates under three modes, namely low-, medium-, and high-security transmission. For each mode, the system is designed according to the corresponding performance metric defined in Section~II-B. Accordingly, the unified optimization problem is formulated as
\begin{subequations}
\label{eq:WG_problem}
\begin{align}
\max_{\mathbf w,\,\boldsymbol\psi^p}
\quad &R^{\mathrm{mode}},
\label{eq:WG_obj}\\
\mathrm{s.t.}\quad
&\sum_{b\in\mathcal B}\left\|\mathbf w_b\right\|_2^2
+\left\|\mathbf w_{\mathrm{AN}}\right\|_2^2
\leq P,
\label{eq:WG_pow}\\
&R^{\mathrm{mode}}\geq R_{\min},
\label{eq:mode_min_constraint}\\
&0\leq x_{n,m}\leq D_x,
\label{eq:WG_pos1}\\
&x_{n,m+1}-x_{n,m}\geq\Delta_{\min}.
\label{eq:WG_pos2}
\end{align}
\end{subequations}

where $R^{\mathrm{mode}}\in\{R^{\mathrm{LST}},R^{\mathrm{MST}},R^{\mathrm{HST}}\}$ is selected according to the required security level. Problem~\eqref{eq:WG_problem} jointly optimizes the transmit beamforming matrix $\mathbf w$ and the PA position set $\boldsymbol\psi^p$ to maximize the mode-dependent system performance under heterogeneous security requirements. Constraint~\eqref{eq:WG_pow} ensures that the total transmit power, including both information signals and AN, does not exceed the available power budget. Constraint~\eqref{eq:mode_min_constraint} guarantees a minimum performance requirement under the selected transmission mode, thereby ensuring a desired level of service quality or secrecy. Constraints~\eqref{eq:WG_pos1} and~\eqref{eq:WG_pos2} enforce the physical deployment limitations of the PASS architecture, including the feasible region of PA locations and the minimum inter-element spacing.

\section{Heterogeneous Security-Aware PPO Design}
\label{sec:hsppo}

In the considered security-aware PASS system, the transmission decision must adapt to the time-varying channel conditions and service security requirements. At each time slot, the security mode determines the authorization role of each user and the corresponding rate or secrecy metric, while the controller jointly configures the information beamforming, AN, and PA positions. Moreover, PA relocation is constrained by the previous configuration, causing the current decision to affect both the instantaneous transmission performance and the feasible decisions in subsequent time slots. Therefore, the proposed joint control problem is formulated as a long-horizon Markov decision process (MDP) with continuous state and action spaces, aiming to optimize the accumulated system utility under mode-dependent transmission requirements. Policy-based DRL is well suited to this problem because it can directly learn continuous control policies without repeatedly solving the nonconvex optimization problem.

In particular, PPO offers a favorable balance between computational complexity and training stability by constraining the magnitude of policy updates through its clipped objective. However, directly applying conventional PPO to the unified three-mode formulation cannot fully exploit the training samples generated under heterogeneous security requirements. Although all three modes share the same control framework, their samples reflect different performance metrics and violation patterns and therefore provide unequal learning values for policy updates. Treating these samples equally may cause samples with larger performance gaps to dominate the shared policy update while weakening informative feedback from other modes. Motivated by this issue, we evaluate the learning value of each sample according to its normalized mode-specific violation severity and adjust its contribution to policy optimization. This design enables more effective cross-mode sample utilization while preserving the low-complexity policy structure and stable update mechanism of PPO.

\subsection{Heterogeneous Security-Aware MDP Formulation}
\label{subsec:security_aware_mdp}

The proposed HSPPO algorithm is adopted to solve the PASS-enhanced security-aware PLS system based on the DRL framework. At each time step $t$, the agent observes the system state $s_t$, samples an action $a_t$ from the stochastic policy $\pi(a_t\mid s_t)$, transitions to the next state $s_{t+1}$ according to the environment dynamics, and receives an instantaneous reward $r_t$. The state, action, and reward are defined as follows:

\textbf{State Space:} The state is designed to capture both the instantaneous propagation conditions and the security context governing the system objective. Specifically, it includes the PA positions from the previous time slot to account for reconfiguration overhead and temporal coupling. It also incorporates the user locations, which determine the geometry-dependent propagation characteristics. In addition, the state incorporates the global CSI, including the effective free-space channels $\mathbf H_{k,t}^{\mathsf H}$, which determine the received signal quality at both legitimate and potential Eves. The security demand information (i.e., low-, medium-, or high-security transmission) is also included to enable adaptive policy learning under heterogeneous security demands. Accordingly, the state at time slot $t$ is defined as
\begin{equation}
 s_t
 \triangleq
 \left\{
 \boldsymbol{\psi}_{t-1}^{p},
 \boldsymbol{\psi}_{k,t},
 \mathbf H_{k,t}^{\mathsf H},
 \mathrm{mode}_t
 \right\}.
 \label{eq:hsppo_state}
\end{equation}
Here, $\boldsymbol{\psi}_{t-1}^{p}$ denotes the PA positions inherited from the previous slot, $\boldsymbol{\psi}_{k,t}$ denotes all user positions in the current slot, $\mathbf H_{k,t}^{\mathsf H}$ represents the downlink channel from the PASS aperture to receiver $k$, and $\mathrm{mode}_t$ specifies the current security requirement.

\textbf{Action Space:} The action consists of continuous control variables that jointly determine transmit beamforming and PA positioning:
\begin{equation}
 \mathbf a_t
 =
 \left\{
 \mathbf w_t,
 \boldsymbol{\psi}_{t}^{p}
 \right\},\quad \forall t,
 \label{eq:hsppo_action}
\end{equation}
where $\mathbf w_t$ denotes the transmit beamforming matrix (including potential AN components), and $\boldsymbol{\psi}_{t}^{p}$ represents the PA position set at time slot $t$. This action directly determines the effective channel $\mathbf H_k^{\mathsf H}\mathbf G$ and thus governs the received SINR and achievable rate/secrecy performance.

\textbf{Reward Function:} The reward function is written in a unified form as
\begin{equation}
r_t^{\mathrm{mode}}
=
R_t^{\mathrm{mode}}
-
\lambda_{\mathrm{mov}}
\left\|
\boldsymbol{\psi}_{t}^{p}
-
\boldsymbol{\psi}_{t-1}^{p}
\right\|_{1}.
\label{eq:reward_all_modes}
\end{equation}
where $r_t^{\mathrm{mode}}$ denotes the immediate reward under the active
security mode, and $R_t^{\mathrm{mode}}$ is the corresponding communication or
secrecy-performance metric. The coefficient $\lambda_{\mathrm{mov}}$ weights the PA
reconfiguration cost, while
$\left\|\boldsymbol{\psi}_{t}^{p}-\boldsymbol{\psi}_{t-1}^{p}\right\|_{1}$ measures the
total PA displacement between two consecutive time slots. Hence, the reward
balances mode-dependent transmission performance and PA-movement overhead.

\subsection{Heterogeneous Security-Aware Policy Optimization}
\label{subsec:hsppo_policy_update}

During training, the agent interacts with the security-level-aware PASS
environment using the current policy and stores the generated transitions in
an on-policy buffer. In addition to the state, action, reward, and next state,
each transition records the active security mode and its corresponding
mode-dependent security performance $R_t^{\mathrm{mode}}$. After a mini-batch
of transitions has been collected, the reward advantage $\hat A_t^{r}$ is
estimated using generalized advantage estimation (GAE). This reward advantage
indicates whether the selected action yields a higher long-term return than
the expected performance under the current state.

However, the reward advantage alone cannot fully characterize the security
quality of a training sample. The unified policy is trained using samples
collected under three security modes with different transmission requirements
and performance metrics. Consequently, an action with a positive reward
advantage may still violate the requirement of the active security mode. To
address this issue, HSPPO first measures the mode-dependent performance relative
to its required threshold and incorporates the resulting violation information
into the PPO update.

Specifically, for the $q$-th sample in the current mini-batch, the
mode-dependent security-performance margin at time step $t$ is defined as
\begin{equation}
\Delta R_t^{(q)}
=
R_t^{\mathrm{mode},(q)}
-
R_{\mathrm{sec}}^{\mathrm{mode}},
\label{eq:mode_security_margin}
\end{equation}
where $N_b$ denotes the mini-batch size, $q\in\{1,\ldots,N_b\}$ denotes the mini-batch sample index, and
$R_{\mathrm{sec}}^{\mathrm{mode}}$ is a mode-dependent required performance
threshold that is treated as a constant for the corresponding security mode.
Accordingly, $\Delta R_t^{(q)}>0$ indicates that the requirement is satisfied,
whereas $\Delta R_t^{(q)}\leq 0$ is counted as a violation.

To quantify the violation severity, the margin is mapped to a smooth
non-negative violation cost as
\begin{equation}
c_t^{\mathrm{mode},(q)}
=
\operatorname{softplus}\!\left(
-\frac{\Delta R_t^{(q)}}{\delta}
\right),
\label{eq:mode_violation_cost}
\end{equation}
where $\delta>0$ is a scaling factor, and
$\operatorname{softplus}(x)=\ln(1+\exp(x))$ is a smooth non-negative function
used to transform the security-performance margin into a differentiable
violation cost.

Since the three security modes may have different numerical ranges, directly
using $c_t^{\mathrm{mode},(q)}$ may cause the policy update to be dominated by
samples with larger violation-cost values. The violation costs are therefore
standardized within each mini-batch as
\begin{equation}
\hat A_t^{\mathrm{vio},(q)}
=
\frac{
c_t^{\mathrm{mode},(q)}-\mu_c
}{
\sigma_c+\varepsilon
},
\label{eq:standardized_violation_cost}
\end{equation}
where $\mu_c$ and $\sigma_c$ denote the mini-batch mean and standard deviation
of the violation costs, respectively, and $\varepsilon>0$ is a small constant
used for numerical stability. This operation transforms the mode-dependent
security violation into a normalized signal that can be combined with the
reward advantage.

To adapt the strength of the violation correction to the current security
status of the policy, the mini-batch violation rate is defined as
\begin{equation}
v
=
\frac{1}{N_b}
\sum_{q=1}^{N_b}
\mathcal{J}\!\left(
\Delta R_t^{(q)}\leq 0
\right),
\label{eq:minibatch_violation_rate}
\end{equation}
where $N_b$ denotes the mini-batch size, $q$ indexes the samples in the current
mini-batch, and $\mathcal{J}(\cdot)$ is the indicator function, which equals one if the enclosed condition holds and zero otherwise. Here, the
subscript $t$ consistently denotes the time-step index, while the superscript
$(q)$ identifies the corresponding mini-batch sample. Based on $v$, an
adaptive security coefficient is calculated as
\begin{equation}
\lambda_{\mathrm{sec}}
=
\sigma\!\left(
\frac{v-v_{\mathrm{tar}}}{\chi}
\right),
\label{eq:adaptive_security_coefficient}
\end{equation}
where $v_{\mathrm{tar}}$ is the target violation rate, $\chi>0$ controls the
smoothness of the adaptation, and
$\sigma(x)=1/(1+\exp(-x))$ denotes the sigmoid function. When the observed
violation rate exceeds $v_{\mathrm{tar}}$, $\lambda_{\mathrm{sec}}$ increases
and strengthens the violation correction; otherwise, the correction is
reduced.

The security-aware advantage of the $q$-th sample is then calculated as
\begin{equation}
\hat A_t^{\mathrm{SA},(q)}
=
\hat A_t^{r,(q)}
-
\lambda_{\mathrm{sec}}
\hat A_t^{\mathrm{vio},(q)}.
\label{eq:smp_advantage}
\end{equation}
When an action achieves a high reward but causes a severe violation, its
positive advantage is reduced or may even be reversed. The extent of this
correction is further adjusted according to the mini-batch violation rate.
Therefore, $\hat A_t^{\mathrm{SA},(q)}$ enables the shared policy to account
for the long-term system reward, the severity of the current sample's
violation, and the overall security status of the policy update.

In addition to violation severity, HSPPO considers whether the reward
advantage evaluates the selected action consistently with the
security-performance margin. A sample is assigned a larger weight when
$\hat A_t^{r,(q)}$ and $\Delta R_t^{(q)}$ provide directionally consistent
evaluations; otherwise, its contribution is reduced to prevent conflicting
reward and security feedback from excessively affecting the shared policy.
The normalized security-consistency weight assigned to the $q$-th sample is
defined as
\begin{equation}
w_t^{\mathrm{SC},(q)}
=
\frac{
1+g\!\left(
\Delta R_t^{(q)},
\hat A_t^{r,(q)}
\right)
}{
\displaystyle
\frac{1}{N_b}
\sum_{j=1}^{N_b}
\left[
1+g\!\left(
\Delta R_t^{(j)},
\hat A_t^{r,(j)}
\right)
\right]
},
\label{eq:smp_weight}
\end{equation}
where $q$ identifies the current mini-batch sample and $j$ is the summation
index over all $N_b$ samples. The numerator evaluates the security consistency
of the $q$-th sample, whereas the denominator averages the consistency scores
over the entire mini-batch. The consistency function is defined as
\begin{equation}
g\!\left(
\Delta R_t^{(q)},
\hat A_t^{r,(q)}
\right)
=
\begin{cases}
\sigma\bigl(
\tfrac{\Delta R_t^{(q)}}{\delta}
\bigr),
& \hat A_t^{r,(q)}\geq 0,\\[0.45em]
\sigma\bigl(
-\tfrac{\Delta R_t^{(q)}}{\delta}
\bigr),
& \hat A_t^{r,(q)}<0.
\end{cases}
\label{eq:margin_consistency_function}
\end{equation}
For a non-negative reward advantage, the consistency score increases with a
positive security-performance margin; for a negative reward advantage, it
increases with a negative margin. The denominator in~\eqref{eq:smp_weight}
normalizes the weights to have unit mean within each mini-batch, thereby
maintaining a stable scale for the policy objective. Inconsistent samples are
not removed from training and are only assigned smaller relative
contributions.

Finally, the probability ratio between the current and old policies for the
$q$-th sample is given by
\begin{equation}
\varrho_t^{(q)}(\theta)
=
\frac{
\pi_\theta\!\left(\mathbf a_t^{(q)}\mid s_t^{(q)}\right)
}{
\pi_{\theta_{\mathrm{old}}}\!\left(\mathbf a_t^{(q)}\mid s_t^{(q)}\right)
}.
\label{eq:smp_probability_ratio}
\end{equation}
Using the security-aware advantage and the security-consistency weight, the
actor network is updated by minimizing
\begin{equation}
\begin{aligned}
&\mathcal L_{\mathrm{actor}}(\theta)
= -\frac{1}{N_b}\sum_{q=1}^{N_b}
w_t^{\mathrm{SC},(q)}
\min\Big(
\varrho_t^{(q)}(\theta)\hat A_t^{\mathrm{SA},(q)},\\
&\operatorname{clip}\!\left(
\varrho_t^{(q)}(\theta),
1-\epsilon_{\mathrm{clip}},
1+\epsilon_{\mathrm{clip}}
\right)
\hat A_t^{\mathrm{SA},(q)}
\Big).
\end{aligned}
\label{eq:smp_ppo_loss}
\end{equation}
where $\epsilon_{\mathrm{clip}}>0$ denotes the PPO clipping parameter. The
critic network is trained using the conventional value loss, while the actor
network is updated for multiple epochs using the collected on-policy samples.
After the update is completed, the buffer is cleared and a new batch of
transitions is generated using the updated policy.

Compared with conventional PPO, the proposed training process introduces two
modifications. First, the normalized mode-dependent violation information
corrects the reward advantage with a coefficient that adapts to the mini-batch
violation rate, preventing high-reward but security-violating actions from being
improperly encouraged while avoiding an unnecessarily strong correction when
the violation rate is already low. Second, the consistency weight
adjusts each sample's contribution according to the agreement between reward
feedback and security performance. These modifications improve the use of
heterogeneous samples generated under the three security modes while
retaining the clipped update and low-complexity policy structure of PPO.

\section{Multi-Relational Hierarchy-Aware Diffusion Policy Optimization Design}
\label{sec:mrhadpo}

\begin{figure*}[!t]
    \centering
    \includegraphics[width=0.80\textwidth]{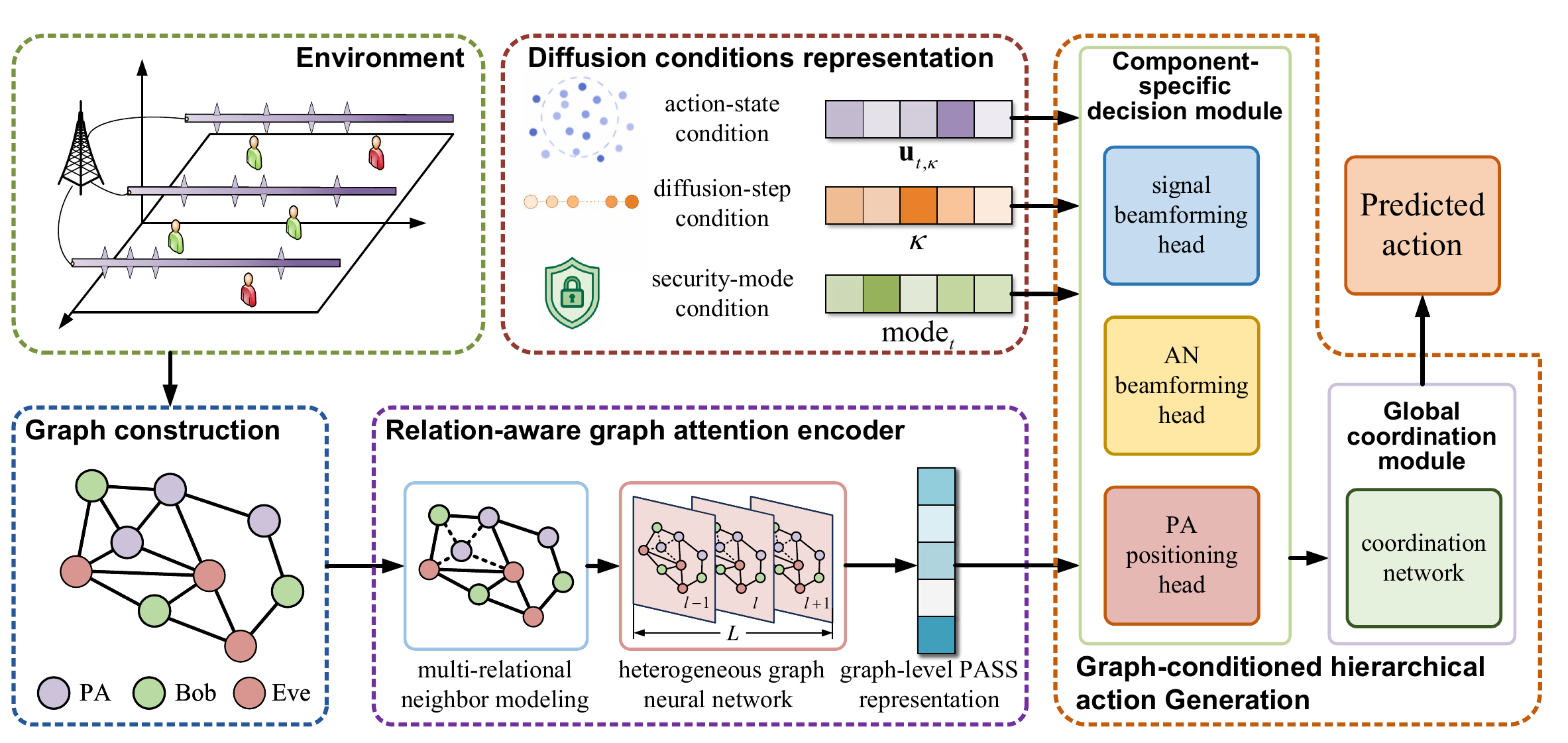}
    \caption{Overall architecture of the proposed MRHA-DPO algorithm.}
    \label{fig:mrha_dpo_architecture}
 
\end{figure*}

Although HSPPO provides a stable and efficient solution for security-aware PASS transmission with heterogeneous security requirements, enhancing policy representation is beneficial for secure control scenarios. In particular, PASS-enabled secure transmission involves intricate interactions among physical deployment, wireless propagation, and security relationships, as well as coupled control variables including beamforming,AN, and PA positions. These characteristics motivate a more expressive policy representation that can exploit the inherent structure of PASS and capture the action generation process.

To this end, we develop an MRHA-DPO algorithm that integrates PASS-specific state representation and structured action generation into diffusion-based policy learning~\cite{ding2025genpo}, as illustrated in Fig.~\ref{fig:mrha_dpo_architecture}. Specifically, a multi-relational graph representation is constructed to characterize the heterogeneous interactions among PAs, legitimate users, and potential eavesdroppers, thereby providing structured state embeddings for policy learning. Based on these representations, a hierarchy-aware action generation mechanism is designed to model the coupled evolution of heterogeneous control variables. The resulting policy is finally optimized through a diffusion policy optimization (DPO) process to achieve effective and adaptive PASS control.

\subsection{Multi-Relational PASS State Representation}
\label{subsec:mr_pass_encoder}

In PASS-enabled secure transmission, the system state exhibits an inherent relational structure rather than being a simple collection of independent variables. The communication quality of each receiver depends jointly on its geometric relationship with the PAs and the resulting wireless channels, while the active security mode further changes the roles of different receivers in determining information leakage. Moreover, the spatial configurations of different PAs are mutually constrained. Therefore, directly flattening the locations, channel states, and security information into a single vector may obscure the heterogeneous dependencies among PAs, legitimate users, and eavesdroppers. To explicitly preserve such structural information, we represent the instantaneous PASS state as a multi-relational heterogeneous graph.

Specifically, the graph at time slot $t$ is defined as
\begin{equation}
\mathcal G_t=
\big(\mathcal V_t, \mathcal E_t, \mathcal R\big),
\label{eq:mr_graph_definition}
\end{equation}
where $\mathcal V_t$ and $\mathcal E_t$ denote the node and directed-edge sets, respectively. $\mathcal R$ is the set of relation types associated with the edges. The node set consists of three types of physical entities,
\begin{align}
\mathcal V_t
&=
\mathcal V^\mathrm{PA}_{t}
\cup
\mathcal V^\mathrm{Bob}_{t}
\cup
\mathcal V^\mathrm{Eve}_{t},
\end{align}
corresponding to the PAs, legitimate receivers, and external eavesdroppers.

To provide a unified input space for heterogeneous nodes, each node $v_{i,t} \in \mathcal V_t$ is represented by a feature vector $x_{i,t}$, which is denoted by
\begin{align}
\mathbf x_{i,t} = \{\boldsymbol\psi_{i,t}, \mathbf o_{i,t}, \mathbf c_{i,t}, \mathrm{mode}_{i,t}, \mathbf g_{i,t}\},
\end{align}
where $\boldsymbol\psi_{i,t}$ denotes the spatial coordinates. $\mathbf{o}_{i,t}$ denotes the one-hot encoded node category, specifying the node category among PAs, legitimate users, and eavesdroppers. $\mathbf{c}_{i,t}$ characterizes the wireless-channel state associated with the node. For PA nodes, $\mathbf{c}_{i,t}$ summarizes the channel characteristics of the corresponding waveguide toward the Bob and Eve sets, including channel magnitude, real and imaginary components, phase descriptors, and the relative Bob/Eve link strengths. For a Bob or Eve node, $\mathbf{c}_{i,t}$ instead describes the channel observed by that receiver from the PASS aperture together with its corresponding link-strength indicators. $\mathrm{mode}_{i,t}$ represents  the active security mode information, which encodes the current low-, medium-, or high-security mode. $\mathbf{g}_{i,t}$ incorporates local geometric information such as the distances to the nearest Bob and Eve. Features that are not applicable to a particular node type are set to zero, allowing all three node categories to share the same input dimension.

The directed edges characterize different interaction semantics among the heterogeneous entities. For each edge $e(i,j)\in \mathcal{E}_{i,j}$, we associate a relation label $r_{i,j} \in \mathcal R$. The $\mathcal R$ is defined as
\begin{align}
\mathcal R = \{r_\text{PB}, r_\text{PE}, r_\text{BE}, r_\text{PP}, r_\text{BB}, r_\text{EE}\},
\end{align}
where $r_\text{PB},r_\text{PE},r_\text{BE},r_\text{PP},r_\text{BB}$ and $r_\text{EE}$ denote the PA$-$Bob, PA$-$Eve, Bob$-$Eve, PA$-$PA, Bob$-$Bob, and Eve$-$Eve relations, respectively. These relations distinguish the heterogeneous interactions involved in secure PASS transmission. In particular, the $r_\text{PB}$ and $r_\text{PE}$ relations characterize the propagation dependencies between the reconfigurable aperture and different receiver roles, the $r_\text{BE}$, $r_\text{BB}$ and $r_\text{EE}$ intra-receiver relations capture security- and interference-related interactions among receivers, and the $r_\text{PP}$ relation reflects the structural coupling among PAs. Importantly, the relation label $r_\text{ij}$ is not treated as an additional continuous edge feature; instead, it determines the relation-specific transformations used during graph attention. Consequently, messages propagated through different types of interactions are processed using different learnable projections, enabling the encoder to distinguish heterogeneous PASS relations during state aggregation.

Given the constructed multi-relational graph, we develop a relation-aware graph attention encoder to extract structured representations of the PASS state. Different from conventional graph attention mechanisms that treat all neighboring interactions with a shared feature transformation, the proposed encoder explicitly
distinguishes heterogeneous relations among PASS entities. Specifically, each graph attention layer is implemented by a heterogeneous graph neural network (HGNN), where relation-specific projections and
interaction-aware priors are jointly incorporated to capture the diverse physical and security dependencies.

\begin{figure}[!t]
\centering
\includegraphics[width=0.86\linewidth]{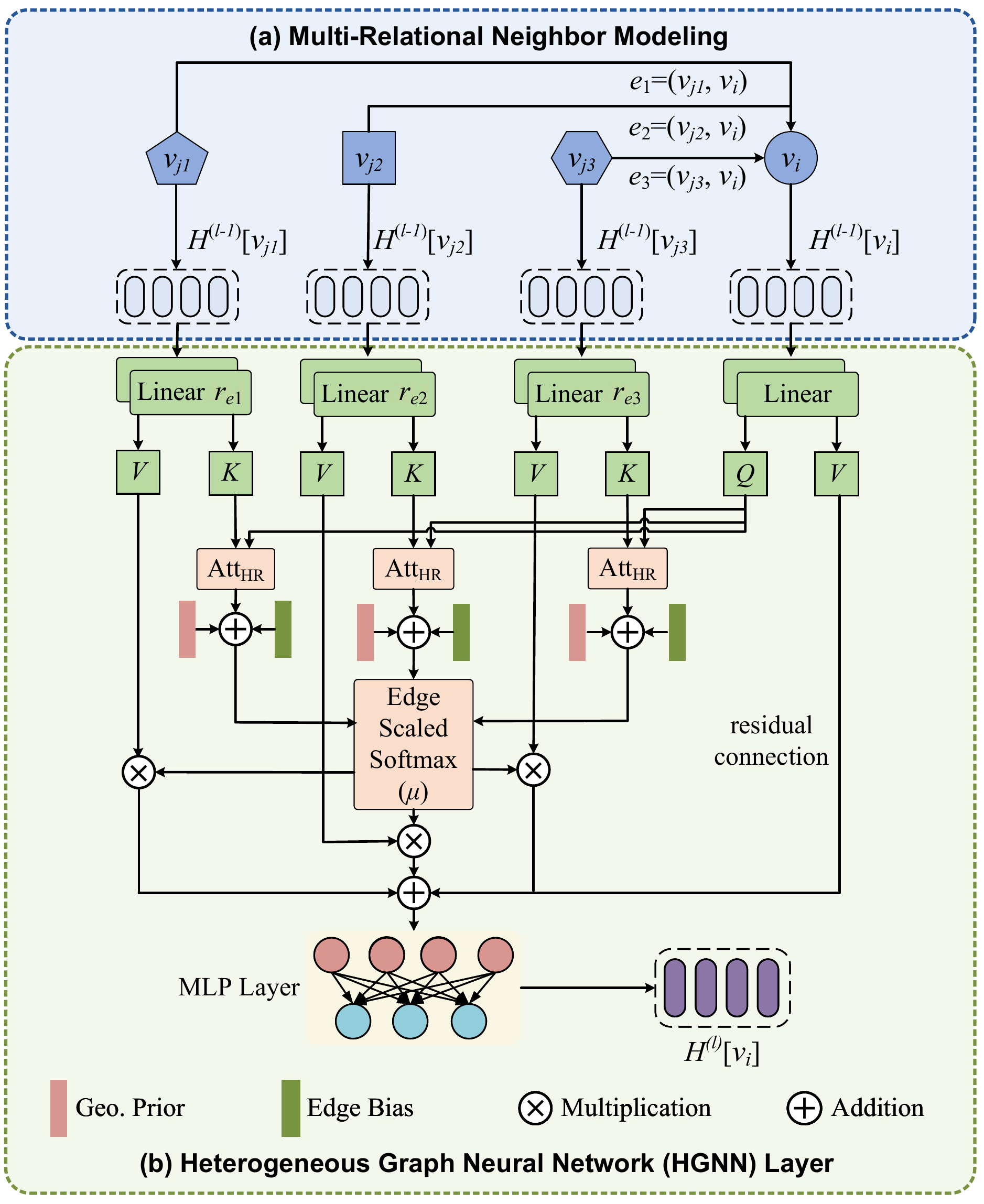}
\caption{Relation-aware graph attention for PASS.}
\label{fig:mrha_hgnn_layer}
\end{figure}

As illustrated in Fig.~\ref{fig:mrha_hgnn_layer}, an HGNN layer performs relation-aware message aggregation for a target node based on its incoming relational edges. For a target node $i$, the neighboring nodes $j\in\mathcal{N}(i)$ and their associated relations are first arranged according to a fixed edge ordering. The objective of the HGNN layer is to adaptively aggregate the information from heterogeneous neighboring
entities while preserving the intrinsic characteristics of the target node.

Let $\mathbf{h}_{i,t}^{(\ell)}$ denote the hidden representation of node $i$ in the $\ell$-th HGNN layer, where the initial representation is given by
\begin{equation}
\mathbf{h}_{i,t}^{(0)}
=
\mathbf{x}_{i,t}.
\end{equation}
For each incoming edge $e(i,j)$ with $j\in\mathcal{N}(i)$ and relation type $r_{ij}\in\mathcal{R}$, the query, key, and value representations are projected according to their corresponding relational semantics.
Specifically, the relation-aware projections are stacked following the ordering of incoming edges:
\begin{align}
\mathbf{Q}_{i,t}^{(\ell)}
&=
\left[
\left(
\mathbf{W}_{r_{ij}}^{Q,(\ell)}
\mathbf{h}_{j,t}^{(\ell-1)}
\right)^{\mathsf T}
\right]_{j\in\mathcal{N}(i)},\\
\mathbf{K}_{i,t}^{(\ell)}
&=
\left[
\left(
\mathbf{W}_{r_{ij}}^{K,(\ell)}
\mathbf{h}_{j,t}^{(\ell-1)}
\right)^{\mathsf T}
\right]_{j\in\mathcal{N}(i)},\\
\mathbf{V}_{i,t}^{(\ell)}
&=
\left[
\left(
\mathbf{W}_{r_{ij}}^{V,(\ell)}
\mathbf{h}_{j,t}^{(\ell-1)}
\right)^{\mathsf T}
\right]_{j\in\mathcal{N}(i)} ,
\end{align}
where $\mathbf{W}_{r_{ij}}^{Q,(\ell)}$, $\mathbf{W}_{r_{ij}}^{K,(\ell)}$, and $\mathbf{W}_{r_{ij}}^{V,(\ell)}$ are learnable projection matrices associated with relation type $r_{ij}$. Each row of $\mathbf{Q}_{i,t}^{(\ell)}$, $\mathbf{K}_{i,t}^{(\ell)}$, and $\mathbf{V}_{i,t}^{(\ell)}$ corresponds to the same incoming edge and its relation semantics.

Based on the above relational projections, we define the prior-enhanced heterogeneous relational attention operator as
\begin{align}
&\operatorname{Att}_{\mathrm{HR}}
\left(
\mathbf Q,\mathbf K,\mathbf V;
\mathbf b^{\mathrm{geo}},
\mathbf b^{\mathrm{edge}}
\right)
\notag\\
&\quad
\triangleq
\operatorname{Softmax}
\left(
\frac{
\left|
\operatorname{diag}
\left(
\mathbf Q\mathbf K^{\mathsf T}
\right)
\right|
}
{\sqrt{d_g}}
+
\mathbf b^{\mathrm{geo}}
+
\mathbf b^{\mathrm{edge}}
\right)^{\mathsf T}
\mathbf V ,
\end{align}
where $d_g$ denotes the dimension of the projected feature space.
The diagonal operation extracts the query--key compatibility of each
incoming edge according to the predefined edge ordering, avoiding
irrelevant pairwise comparisons among different neighbors. Moreover,
$\mathbf b^{\mathrm{geo}}$ and $\mathbf b^{\mathrm{edge}}$ introduce
additional physical and semantic priors, where
$\mathbf b^{\mathrm{geo}}$ characterizes the geometric and propagation
conditions, while $\mathbf b^{\mathrm{edge}}$ encodes relation-dependent
edge information and active security states.

The aggregated relational representation for node $i$ is then obtained as
\begin{equation}
\mathbf m_{i,t}^{(\ell)}
=
\operatorname{Att}_{\mathrm{HR}}
\left(
\mathbf{Q}_{i,t}^{(\ell)},
\mathbf{K}_{i,t}^{(\ell)},
\mathbf{V}_{i,t}^{(\ell)};
\mathbf b_{i,t}^{\mathrm{geo},(\ell)},
\mathbf b_{i,t}^{\mathrm{edge},(\ell)}
\right).
\end{equation}
The message $\mathbf m_{i,t}^{(\ell)}$ aggregates features from heterogeneous neighbors, where attention weights are jointly determined by relational feature compatibility, physical propagation characteristics, edge semantics, and security requirements.

To preserve the intrinsic state information of the target node, a relation-independent self representation is further introduced, while a multilayer perceptron (MLP) is employed to fuse the self representation with the aggregated relational information:
\begin{align}
\mathbf s_{i,t}^{(\ell)}
&=
\mathbf W_{\mathrm{self}}^{(\ell)}
\mathbf h_{i,t}^{(\ell-1)},\\
\mathbf h_{i,t}^{(\ell)}
&=
\operatorname{MLP}^{(\ell)}
\left(
\mathbf s_{i,t}^{(\ell)}
+
\mathbf m_{i,t}^{(\ell)}
\right).
\end{align}
Here, $\mathbf{s}_{i,t}^{(\ell)}$ preserves the intrinsic information of the target node, while $\mathbf{m}_{i,t}^{(\ell)}$ captures the relation-aware context aggregated from neighboring entities. The MLP performs nonlinear feature transformation and fusion to obtain the updated node representation. By stacking multiple HGNN layers, higher-order dependencies among PAs, legitimate users, and eavesdroppers can be progressively captured.

After the final HGNN layer, the obtained node representations are further transformed into a graph-level PASS state embedding through a role-aware pooling strategy. Specifically, node representations corresponding to different physical roles $\rho\in \{\mathrm{PA},\mathrm{Bob},\mathrm{Eve}\}$ are separately aggregated as
\begin{equation}
\mathbf h_{\rho,t}^{\mathrm{RP}}
=
\operatorname{AttPool}_{\rho}
\left(
\left\{
\mathbf h_{i,t}^{(L_g)}
\mid i\in\mathcal V_{\rho,t}
\right\}
\right).
\end{equation}
Meanwhile, the global graph-level information is summarized through average pooling:
\begin{equation}
\mathbf h_t^{\mathrm{GAP}}
=
\frac{1}{|\mathcal V_t|}
\sum_{i\in\mathcal V_t}
\mathbf h_{i,t}^{(L_g)}.
\end{equation}
Finally, the structured PASS state representation is obtained as
\begin{equation}
\mathbf z_t^G
=
\operatorname{MLP}
\left(
\mathbf h_{\mathrm{PA},t}^{\mathrm{RP}};
\mathbf h_{\mathrm{Bob},t}^{\mathrm{RP}};
\mathbf h_{\mathrm{Eve},t}^{\mathrm{RP}};
\mathbf h_t^{\mathrm{GAP}}
\right),
\end{equation}
which preserves role-specific physical characteristics while capturing the global PASS topology, providing a structured state representation for subsequent diffusion-based policy learning.

\subsection{Graph-Conditioned Hierarchy-Aware Action Generator}
\label{subsec:mrha_velocity_network}

The structured state representation obtained above captures the
heterogeneous physical and security dependencies of the PASS system.
However, translating such structured state information into effective
control decisions remains nontrivial, since the action space consists
of multiple heterogeneous components with distinct physical meanings
and constraints. In particular, transmission-related variables and PA
configuration variables operate in different domains, while their
effects on communication reliability and security are strongly coupled.
Directly generating all action dimensions through a single flattened
mapping may obscure such structural differences and make it difficult
to coordinate the coupled decisions.

To address this issue, we introduce a hierarchy-aware action generator
that explicitly separates component-specific decision modeling from
cross-component coordination. At the lower level, dedicated branches
characterize the evolution of different action components according to
their respective structures and constraints. At the higher level, their
intermediate representations are jointly coordinated to capture the
coupling among heterogeneous decisions. In this manner, the generator
preserves variable-specific characteristics while maintaining global
consistency across the joint PASS control space.

To realize the hierarchical generation mechanism, we parameterize the evolution of the intermediate action representation through a graph-conditioned velocity field. Let $\mathbf u_{t,\kappa}$ denote a generic intermediate variable at normalized diffusion time $\kappa$. Its specific construction and evolution within the reversible diffusion policy will be introduced in the next subsection. Here, we focus on how the hierarchy-aware generator models its evolution direction according to the heterogeneous PASS action structure. Specifically, three component-specific heads first generate preliminary velocity components for signal beamforming, AN beamforming, and PA positioning:

\begin{equation}
\begin{aligned}
\dot{\mathbf w}_{t,\kappa}
&=
\operatorname{Head}_{\mathrm{BF}}\!\left(
\mathbf z_t^{G},\mathrm{mode}_t,
\mathbf u_{t,\kappa},\kappa
\right),\\
\dot{\boldsymbol\beta}_{t,\kappa}
&=
\operatorname{Head}_{\mathrm{AN}}\!\left(
\mathbf z_t^{G},\mathrm{mode}_t,
\mathbf u_{t,\kappa},\kappa
\right),\\
\dot{\boldsymbol\alpha}_{t,\kappa}
&=
\operatorname{Head}_{\mathrm{POS}}\!\left(
\mathbf z_t^{G},\mathrm{mode}_t,
\mathbf u_{t,\kappa},\kappa
\right).
\end{aligned}
\label{eq:hdcd_action_heads}
\end{equation}
The three velocity components retain the distinct physical structures of the corresponding action variables.

To account for their coupled effects on transmission rate, information leakage, and PASS geometry, the preliminary components are jointly processed by a global coordination module:
\begin{equation}
\mathbf h_{t,\mathrm{coord}}^{\kappa}
=
\Phi\!\left(
\dot{\mathbf w}_{t,\kappa},
\dot{\boldsymbol\beta}_{t,\kappa},
\dot{\boldsymbol\alpha}_{t,\kappa},
\mathbf z_t^{G},
\mathrm{mode}_t,
\mathbf u_{t,\kappa},
\kappa
\right).
\label{eq:hdcd_coordination}
\end{equation}
The coordinated velocity components are then obtained as
\begin{equation}
\left(
\dot{\mathbf w}_{t,\kappa}^{*},
\dot{\boldsymbol\beta}_{t,\kappa}^{*},
\dot{\boldsymbol\alpha}_{t,\kappa}^{*}
\right)
=
\Psi\!\left(
\mathbf h_{t,\mathrm{coord}}^{\kappa}
\right).
\label{eq:hdcd_refinement}
\end{equation}
Their concatenation defines the complete conditional velocity field:
\begin{equation}
\mathbf v_{\omega}\!\left(
\mathbf u_{t,\kappa},
\mathbf z_t^{G},
\mathrm{mode}_t,
\kappa
\right)
\triangleq
\left[
\left(\dot{\mathbf w}_{t,\kappa}^{*}\right)^{\mathsf T},\!
\left(\dot{\boldsymbol\beta}_{t,\kappa}^{*}\right)^{\mathsf T},\!
\left(\dot{\boldsymbol\alpha}_{t,\kappa}^{*}\right)^{\mathsf T}
\right]^{\mathsf T}.
\label{eq:mrha_velocity_field}
\end{equation}
where $\omega$ denotes the parameters associated with the hierarchical
action generator. In implementation, the real and imaginary parts of
the information and AN beamformers are represented by real-valued
coordinates during action evolution and are subsequently reconstructed
by the action decoder.

Through the above two-layer design, the lower layer preserves the component-specific characteristics of heterogeneous PASS control variables, while the higher layer coordinates their cross-component dependencies. The resulting conditional velocity field provides the action-evolution model used by the diffusion policy, whose generation and optimization procedure is introduced in the following subsection.

\subsection{Multi-Relational Hierarchy-Aware Diffusion Policy Optimization and Training Process}
\label{subsec:mrha_dpo_training}

With the structured PASS state representation and hierarchy-aware action
generator established above, we next describe how they are integrated
into the complete MRHA-DPO policy and optimized in an end-to-end manner.
In MRHA-DPO, DPO serves as the policy
optimization backbone, providing an expressive generative distribution
for the highly coupled PASS action space. Meanwhile, the reversible
diffusion construction enables exact policy-likelihood evaluation,
which allows the resulting diffusion policy to be directly optimized
within a PPO-style on-policy framework.

Different from conventional PPO, which commonly parameterizes a
continuous policy using a unimodal Gaussian distribution, the adopted
DPO backbone constructs the policy through two coupled augmented-action
variables. Specifically, at each time slot $t$, the initial augmented
action is defined as
\begin{equation}
\tilde{\mathbf a}_{t,0}
=
\left(
\mathbf x_{t,0},
\mathbf y_{t,0}
\right),
\qquad
\mathbf x_{t,0},
\mathbf y_{t,0}
\sim
\mathcal N(\mathbf 0,\mathbf I).
\label{eq:augmented_action_initial}
\end{equation}

Conditioned on the graph representation $\mathbf z_t^{G}$ and the
active security mode $\mathrm{mode}_t$, the hierarchy-aware action
generator introduced in the preceding subsection parameterizes the
conditional velocity field $\mathbf v_{\omega}
\left(
\mathbf u_{t,\kappa},
\mathbf z_t^{G},
\mathrm{mode}_t,
\kappa
\right),
$
where
$\mathbf u_{t,\kappa}\in
\{\mathbf x_{t,\kappa},\mathbf y_{t,\kappa}\}$
denotes either augmented-action variable at normalized diffusion time
$\kappa$. Together with the reversible mixing operation, the velocity
field evolves the two variables over $K_d$ diffusion steps and produces
the terminal augmented action
$
\tilde{\mathbf a}_{t,1}
=
\left(
\mathbf x_{t,1},
\mathbf y_{t,1}
\right).
$

The physical PASS action executed in the environment is obtained from
the two terminal variables as
$
\mathbf a_t
=
\frac{
\mathbf x_{t,1}
+
\mathbf y_{t,1}
}{2}.
$
This construction allows the diffusion policy to explore an augmented
action space while maintaining a unique physical action for interaction
with the PASS environment.

The coupled diffusion evolution defines an invertible conditional
mapping
\begin{equation}
f_{\omega}:
\left(
\mathbf x_{t,0},
\mathbf y_{t,0}
\right)
\mapsto
\left(
\mathbf x_{t,1},
\mathbf y_{t,1}
\right).
\label{eq:invertible_diffusion_mapping}
\end{equation}
Accordingly, the terminal variables can be exactly inverted to recover
their initial Gaussian variables. The corresponding policy
log-likelihood is evaluated through the change-of-variables formula:
\begin{equation}
\log\!\pi_{\omega}
\left(
\tilde{\mathbf a}_{t,1}
\mid
\mathbf s_t
\right)
=
\log\!p_0
\left(
\mathbf x_{t,0},
\mathbf y_{t,0}
\right)
-
\log
\left|
\det
\frac{
\partial f_{\omega}
}{
\partial
\left(
\mathbf x_{t,0},
\mathbf y_{t,0}
\right)
}
\right|,
\label{eq:exact_diffusion_likelihood}
\end{equation}
where $p_0(\cdot)$ denotes the standard Gaussian prior. The exact and
differentiable likelihood in
\eqref{eq:exact_diffusion_likelihood}
avoids heuristic density approximation and enables the diffusion policy
to be optimized using PPO-style importance sampling.

Based on the above reversible diffusion policy, the MRHA-DPO training
procedure consists of three stages: action sampling, exact trajectory
inversion, and policy optimization. During action sampling, the current
policy evolves the initial augmented variables to obtain
$\tilde{\mathbf a}_{t,1}$, and the corresponding physical action
$\mathbf a_t$ is executed
in the environment. The resulting transition, active security mode,
terminal augmented variables, and reward are stored in the on-policy
buffer.

During policy optimization, the reversible mapping is first applied
backward to recover $(\mathbf x_{t,0},\mathbf y_{t,0})$ from the
terminal augmented action. The exact likelihood in
\eqref{eq:exact_diffusion_likelihood} is then evaluated under both the
current and old policies. Accordingly, the diffusion-policy probability
ratio is defined as
\begin{equation}
\rho_t(\omega)
=
\frac{
\pi_{\omega}
\left(
\tilde{\mathbf a}_{t,1}
\mid
\mathbf s_t
\right)
}{
\pi_{\omega_{\mathrm{old}}}
\left(
\tilde{\mathbf a}_{t,1}
\mid
\mathbf s_t
\right)
}.
\label{eq:dpo_policy_ratio}
\end{equation}

Using the above ratio and the GAE advantage $\hat A_t$, the PPO clipped
loss is given by
\begin{equation}
\begin{aligned}
\mathcal L_{\mathrm{CLIP}}(\omega)
=-\mathbb E_t\!\Big[
\min\big(&\rho_t(\omega)\widehat A_t,\\[-0.2em]
&\operatorname{clip}\!\left(\rho_t(\omega),1-\epsilon_{\mathrm{clip}},1+\epsilon_{\mathrm{clip}}\right)\widehat A_t
\big)
\Big].
\end{aligned}
\label{eq:dpo_clip_loss}
\end{equation}

To encourage exploration under the exact diffusion-policy likelihood,
the likelihood-based entropy regularization term is defined as
\begin{equation}
\mathcal L_{\mathrm{EEL}}(\omega)
=
\mathbb E_t
\left[
\rho_t(\omega)
\log
\pi_{\omega}
\left(
\tilde{\mathbf a}_{t,1}
\mid
\mathbf s_t
\right)
\right].
\label{eq:dpo_eel_loss}
\end{equation}

Moreover, since the two terminal augmented variables correspond to the
same physical PASS action, excessive discrepancy between them may lead
to redundant exploration in the augmented action space. We therefore
propose the terminal compression loss
\begin{equation}
\mathcal L_{\mathrm{CMP}}(\omega)
=
\mathbb E_t
\left[
\left\|
\mathbf x_{t,1}
-
\mathbf y_{t,1}
\right\|_2^2
\right].
\label{eq:dpo_compression_loss}
\end{equation}

The final MRHA-DPO actor objective is formulated as
\begin{equation}
\mathcal L_{\mathrm{actor}}^{\mathrm{DPO}}(\omega)
=
\mathcal L_{\mathrm{CLIP}}(\omega)
+
\xi
\mathcal L_{\mathrm{EEL}}(\omega)
+
\zeta
\mathcal L_{\mathrm{CMP}}(\omega),
\label{eq:dpo_actor_loss}
\end{equation}
where $\xi$ and $\zeta$ control the strengths of exploration
regularization and terminal-variable consistency, respectively.

The critic estimates the long-term state value and is trained using
\begin{equation}
\mathcal L_{\mathrm{critic}}(\phi)
=
\mathbb E_t
\left[
\left(
V_{\phi}(\mathbf s_t)
-
\hat R_t
\right)^2
\right],
\label{eq:dpo_critic_loss}
\end{equation}
where $\hat R_t$ denotes the return target. The actor and critic are
updated using the collected on-policy samples. During actor optimization,
gradients are propagated through the reversible diffusion backbone,
hierarchy-aware action generator, and multi-relational PASS state
encoder. Consequently, MRHA-DPO jointly learns the structured state
representation, coupled action-generation mechanism, and long-horizon
value estimator in an end-to-end manner.

\section{Numerical Results}
\label{sec:numerical_results}

This section evaluates the learning behavior, system performance, ablation results, and spatial energy patterns of the proposed security-aware PASS framework. Unless otherwise stated, all methods use the same channel model, power constraint, and PA-feasibility constraints.

\subsection{Simulation Setup}
\label{subsec:simulation_setup}

The simulations are conducted in a $20\times20~\mathrm{m}^2$ service region, where four Bobs and two Eves are distributed on the ground plane. The PASS consists of $N=4$ parallel dielectric waveguides, with $M=4$ PAs deployed on each waveguide at a height of $H=3~\mathrm{m}$. The carrier frequency is set to $f_c=28~\mathrm{GHz}$, and the effective refractive index of the dielectric waveguide is $n_{\mathrm{eff}}=1.4$. The minimum inter-PA spacing is $\Delta_{\min}=\lambda/2$. The maximum transmit power at the BS is $P_{\max}=30~\mathrm{dBm}$, and the receiver noise power is set to $-90~\mathrm{dBm}$ for both Bobs and Eves. 

To characterize the dynamic transmission environment, the locations of Bobs and Eves are updated at each time slot within the considered service region. Meanwhile, the security demand is randomly selected from LST, MST, and HST at each time slot, such that a single policy is trained under heterogeneous and time-varying security requirements. All DRL algorithms are trained for 2000 episodes, with 100 time steps in each episode. A rollout batch size of 1024 is adopted, and the performance is evaluated every 50 episodes over 10 independent runs. For all proposed algorithms, the discount factor, GAE parameter, and clipping parameter are set to $\gamma=0.98$, $\lambda_{\mathrm{GAE}}=0.95$, and $\epsilon_{\mathrm{clip}}=0.2$, respectively.

For HSPPO, both the actor and critic networks contain three hidden layers with 64 neurons per layer. The scaling factor of the smooth violation cost in Eq.~\eqref{eq:mode_violation_cost} is set to $\delta=0.35$, while the smoothness parameter of the adaptive security coefficient in Eq.~\eqref{eq:adaptive_security_coefficient} is set to $\chi=0.08$. For MRHA-DPO, the multi-relational graph encoder contains four HGNN layers with 128 hidden units and eight radial-basis distance features. The component-specific heads and the global coordination network contain 96 and 128 hidden units, respectively. The diffusion process employs $K_d=5$ evolution steps. The entropy-regularization coefficient and terminal-compression coefficient in (58) are set to $\xi=0.0025$ and $\zeta=0.020$, respectively. The learning rates of the actor and critic are $3.4\times10^{-4}$ and $4.2\times10^{-4}$, respectively.

\subsection{Convergence and Learning Performance}
\label{subsec:convergence_performance}

\begin{figure*}[!t]
    \centering
    \subfloat[LST]{%
        \includegraphics[width=0.29\textwidth]{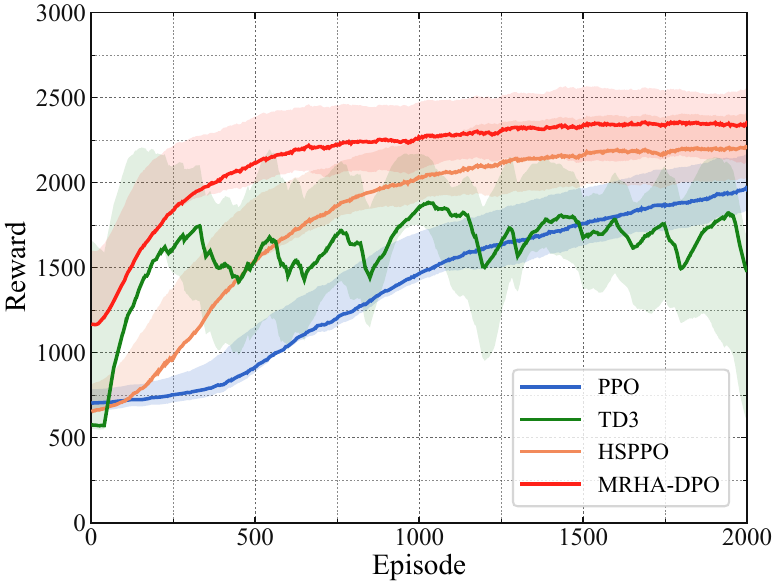}%
        \label{fig:reward_lst}}
    \hfill
    \subfloat[MST]{%
        \includegraphics[width=0.29\textwidth]{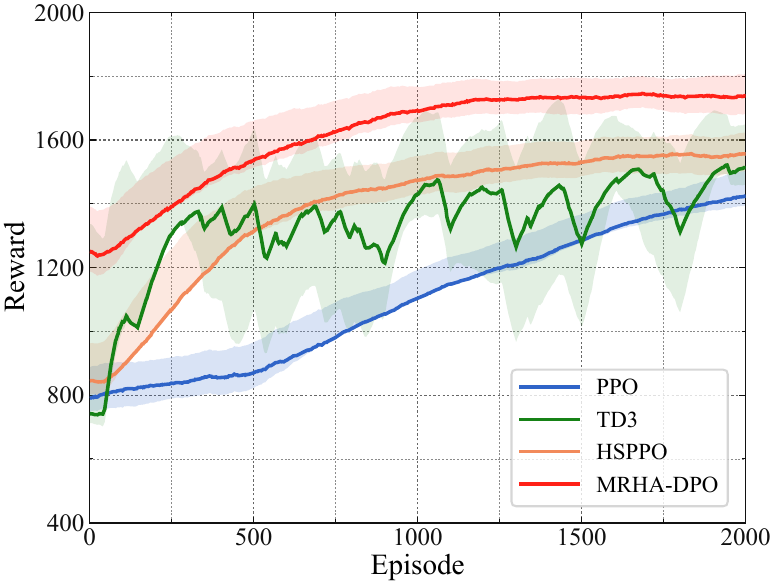}%
        \label{fig:reward_mst}}
    \hfill
    \subfloat[HST]{%
        \includegraphics[width=0.29\textwidth]{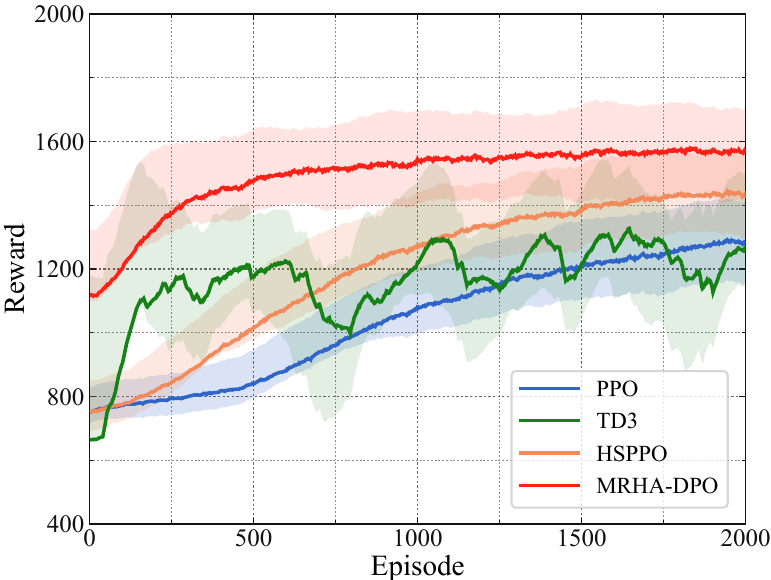}%
        \label{fig:reward_hst}}

    \caption{Training reward under different security demands.}
    \label{fig:training_reward}
\end{figure*}

Fig.~\ref{fig:training_reward} compares the training rewards of different DRL algorithms under the LST, MST, and HST modes. PPO, HSPPO, and MRHA-DPO gradually reach stable reward levels under all three security requirements, demonstrating that the formulated long-horizon control problem can be effectively optimized by policy-based DRL. In contrast, twin delayed deep deterministic policy gradient (TD3) exhibits persistent fluctuations and fails to achieve stable convergence, as its off-policy actor--critic updates are more sensitive to the strongly coupled action space and non-smooth security objectives. Among the convergent methods, both proposed algorithms consistently outperform conventional PPO across the three security modes. HSPPO achieves stable performance improvements by incorporating mode-dependent violation information and security-consistency weighting while retaining a lightweight Gaussian policy structure. MRHA-DPO further enhances the achievable reward through multi-relational PASS representation and diffusion-based action generation, providing stronger modeling capability for heterogeneous relations and coupled actions. Their consistent performance gains under LST, MST, and HST also demonstrate good adaptability to heterogeneous security requirements.

\subsection{Impact of Key System Parameters}
\label{subsec:system_parameter_results}

\subsubsection{Impact of Transmit Power}

\begin{figure}[h]
    \centering
    \includegraphics[width=0.72\columnwidth]{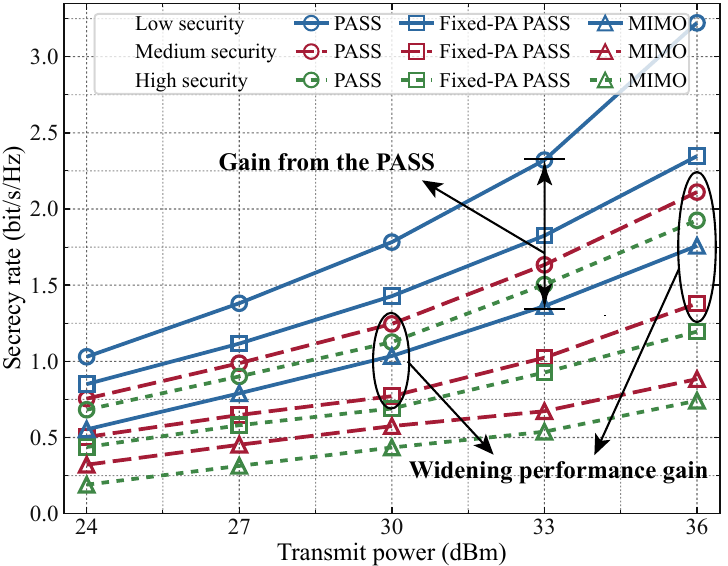}
    \caption{Secrecy rate versus transmit power.}
    \label{fig:exp2_transmit_power}
\end{figure}

Fig.~\ref{fig:exp2_transmit_power} shows the mode-dependent secrecy rate versus the transmit power. Under the same security requirement, the movable-PA PASS consistently achieves the best performance, followed by the fixed-PA PASS and conventional multiple-input multiple-output (MIMO). This is because PASS provides additional spatial degrees of freedom for shaping the effective channels, while movable PAs can further adapt the propagation geometry to the instantaneous user distribution. As the transmit power increases, the performance gain of movable-PA PASS over the other schemes gradually increases. Higher transmit power provides more flexibility for jointly optimizing information beamforming and AN, while PA repositioning enables these transmission resources to be more effectively utilized for desired-signal enhancement and leakage suppression. For the same transmission scheme, LST achieves the highest secrecy rate, followed by MST and HST. LST does not impose secrecy constraints, whereas MST and HST require progressively stronger leakage suppression as the set of potential interceptors increases. The performance gaps among the three security modes also increase with the transmit power, since the additional transmit power can be more directly exploited in LST, while part of the increased transmission capability in MST and HST is used to satisfy the corresponding security requirements.

\subsubsection{Impact of Service-Region Size}

\begin{figure}[h]
    \centering
    \includegraphics[width=0.72\columnwidth]{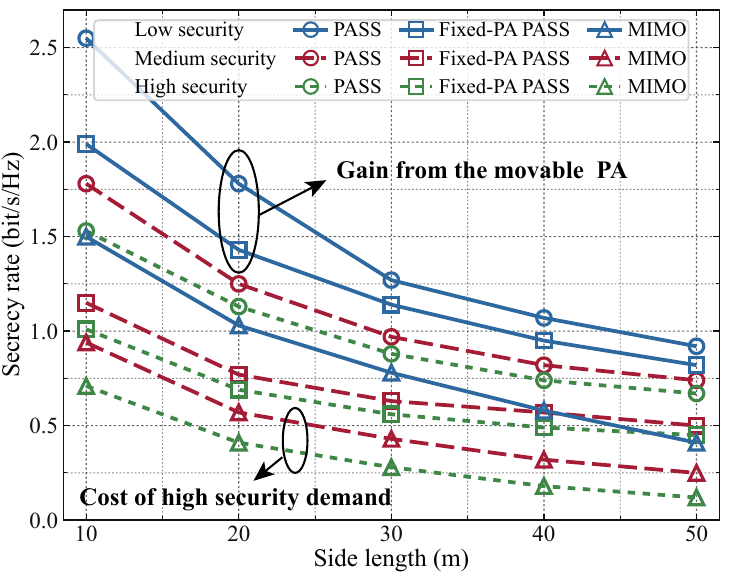}
    \caption{Secrecy rate versus service-region side length.}
    \label{fig:exp3_side_length}
\end{figure}

Fig.~\ref{fig:exp3_side_length} shows the mode-dependent rate performance versus the side length of the service region. The performance of all schemes decreases as the region expands because the increased distances between users and radiating elements result in higher free-space propagation loss. Under each security mode, movable-PA PASS achieves the highest rate, followed by fixed-PA PASS and conventional MIMO. The absolute gap between the two PASS schemes gradually narrows because widely distributed users favor more coverage-balanced PA placements, reducing the marginal benefit of local PA repositioning. Nevertheless, PASS retains a clear advantage over MIMO through guided-wave signal delivery and distributed radiation. For example, under LST, although the absolute movable-PA PASS and MIMO gap decreases from approximately 1.05 to 0.50 bit/s/Hz as the side length increases from 10 to 50 m, the corresponding relative gain increases from approximately 70\% to 119\%, indicating that MIMO degrades more rapidly. Finally, the performance follows the order LST, MST, and HST, while the gaps among these modes narrow as propagation loss gradually dominates the effect of their different security constraints.

\subsection{Ablation Study of the Proposed Algorithms}
\label{subsec:ablation_results}

\begin{table}[h]
\centering
\caption{Ablation results in terms of $R_{\min}$ (bits/s/Hz), where the check mark, cross, and dash denote enabled, ablated, and inapplicable components, respectively.}
\label{tab:combined_ablation}
\scriptsize
\renewcommand{\arraystretch}{0.92}
\setlength{\tabcolsep}{1.10pt}
\begin{tabular}{@{}lccccccc@{}}
\toprule
\textbf{Algorithm}
& \shortstack{\textbf{Advantage}\\\textbf{Correction}}
& \shortstack{\textbf{Consistency}\\\textbf{Weighting}}
& \shortstack{\textbf{Graph}\\\textbf{Encoder}}
& \shortstack{\textbf{Diffusion}\\\textbf{Actor}}
& \textbf{LST}
& \textbf{MST}
& \textbf{HST} \\
\midrule
\multirow{4}{*}{\textbf{HSPPO}}
& \xmark & \xmark & -- & --
& 1.2039 & 0.5795 & 0.5580 \\
& \cmark & \xmark & -- & --
& 1.3686 & 0.7513 & 0.7131 \\
& \xmark & \cmark & -- & --
& 1.3680 & 0.7396 & 0.7108 \\
& \cmark & \cmark & -- & --
& \textbf{1.3804} & \textbf{0.7549} & \textbf{0.7278} \\
\midrule
\multirow{4}{*}{\textbf{MRHA-DPO}}
& -- & -- & \xmark & \xmark
& 1.2039 & 0.5795 & 0.5580 \\
& -- & -- & \cmark & \xmark
& 1.5463 & 1.1302 & 0.9523 \\
& -- & -- & \xmark & \cmark
& 1.4984 & 1.0474 & 0.8711 \\
& -- & -- & \cmark & \cmark
& \textbf{1.7826} & \textbf{1.2457} & \textbf{1.1264} \\
\bottomrule
\end{tabular}
\end{table}

Table~\ref{tab:combined_ablation} evaluates the contributions of the main components in HSPPO and MRHA-DPO under the three security requirements. For HSPPO, either advantage correction or security-consistency weighting improves the minimum rate over conventional PPO, while their joint use achieves the best performance across LST, MST, and HST. The improvement is more evident under MST and HST, indicating that the proposed security-aware policy update is particularly effective when stronger secrecy constraints are imposed.

For MRHA-DPO, both the multi-relational graph encoder and the diffusion policy improve the performance when introduced individually. The graph encoder enhances the representation of heterogeneous PA--user and security relations, while the diffusion policy provides stronger modeling capability for the coupled action space. Combining the two components achieves the highest performance under all three security modes, confirming the effectiveness of jointly exploiting structured state representation and expressive action generation.

\subsection{Spatial Energy Distribution of Security-Aware PASS}
\label{subsec:spatial_results}

\begin{figure}[h]
    \centering
    \includegraphics[width=0.84\columnwidth]{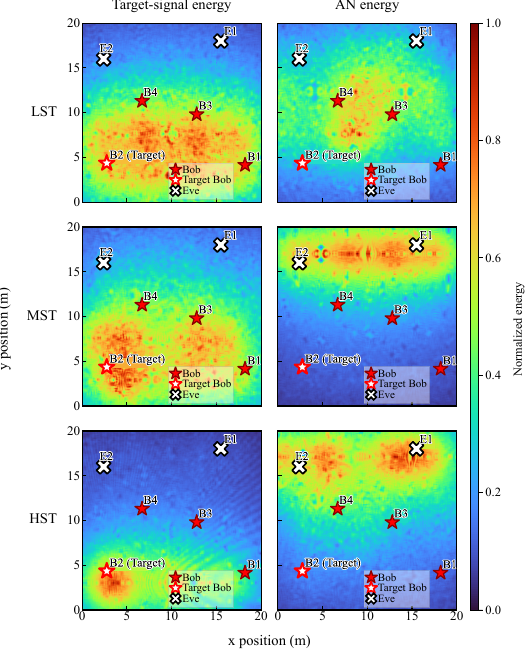}
    \caption{Normalized target-signal and AN energy distributions for Bob 2 under different security modes. The left and right columns show the target-signal and AN energy, respectively, while the top, middle, and bottom rows correspond to LST, MST, and HST, respectively.}
    \label{fig:exp5_spatial_energy}
\end{figure}

Fig.~\ref{fig:exp5_spatial_energy} illustrates the normalized spatial distributions of the target-signal and AN energy under different security requirements. Under LST, where no secrecy constraint is imposed, the transmission strategy mainly concentrates the desired-signal energy around the target Bob, while the spatial distribution of AN is not shaped toward specific unintended receivers. Under MST, AN is generated to suppress information leakage toward external Eves while maintaining the desired signal around the target Bob. Under HST, where both Eves and non-target Bobs are regarded as potential interceptors, the target-signal distribution becomes more concentrated and the AN distribution is accordingly adjusted to provide stronger information-leakage suppression. These distinct spatial energy patterns demonstrate that the proposed scheme can adapt its transmission strategy to different security requirements, further validating the effectiveness of the security-aware PASS design.

\section{Conclusion}

This paper investigated PASS-enabled PLS under heterogeneous confidentiality requirements. A unified framework was established for low-, medium-, and high-security transmission, with joint security-aware beamforming and PA positioning formulated as a long-horizon decision problem. Two complementary algorithms, HSPPO and MRHA-DPO, were developed for lightweight stable control and enhanced optimization, respectively. Simulation results showed that movable PAs consistently outperform fixed-PA PASS and conventional MIMO in secrecy rate, with larger gains at higher transmit power. HSPPO achieved stable cross-mode control with low policy complexity, while MRHA-DPO delivered the best overall performance by exploiting heterogeneous PASS interactions and coupled action structures. These results demonstrate that PASS reconfigurability can be effectively translated into reliability and secrecy gains, while enabling robust adaptation to heterogeneous security requirements.

\balance
\bibliographystyle{IEEEtran}
\bibliography{mybib}

\end{document}